\documentclass[%
 reprint,
superscriptaddress,
 amsmath,amssymb,longbibliography,
 aps,
prl,
]{revtex4-1}
\pdfoutput=1
\usepackage{hyperref}
\usepackage{graphicx}
\usepackage{amsmath}
\usepackage{color}
 
\usepackage{pdfpages} 
\usepackage{pgffor} 
\makeatletter
\AtBeginDocument{\let\LS@rot\@undefined}
\makeatother

\def\supplementfilename{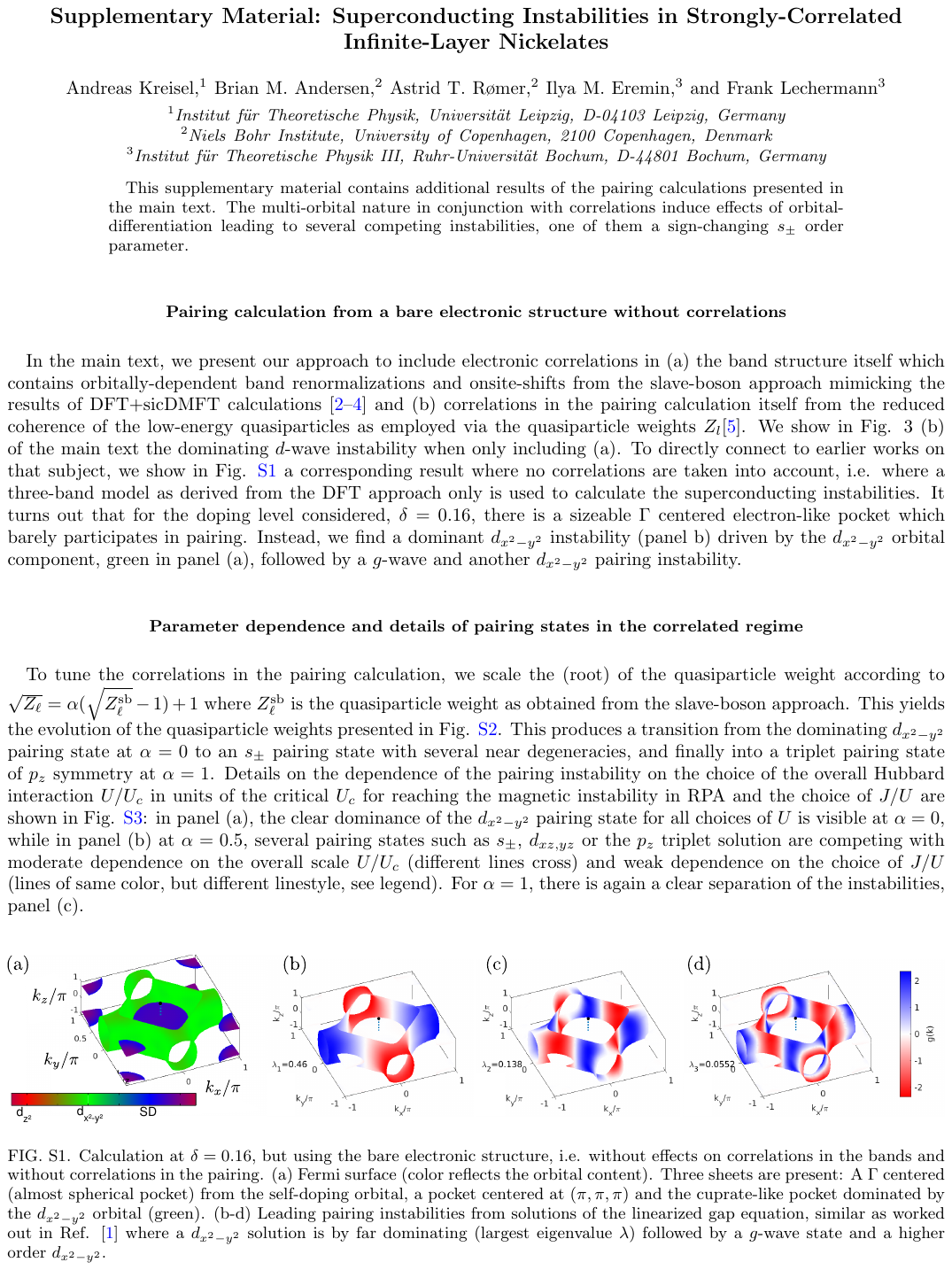}
\pdfximage{\supplementfilename}
\def\numbersupplementpages{\the\pdflastximagepages}
\newif\ifarXiv
\arXivtrue 

\begin{document}
\title{Superconducting Instabilities in Strongly-Correlated Infinite-Layer Nickelates}

\author{Andreas Kreisel}
\affiliation{Institut f\" ur Theoretische Physik, Universit\"at Leipzig, D-04103 Leipzig, Germany}
\author{Brian M. Andersen}
\affiliation{Niels Bohr Institute, University of Copenhagen, 2100 Copenhagen, Denmark}
\author{Astrid T. R{\o}mer}
\affiliation{Niels Bohr Institute, University of Copenhagen, 2100 Copenhagen, Denmark}
\author{Ilya M. Eremin}
\affiliation{Institut f\"ur Theoretische Physik III, Ruhr-Universit\"at Bochum, D-44801 Bochum, Germany}
\author{Frank Lechermann}
\affiliation{Institut f\"ur Theoretische Physik III, Ruhr-Universit\"at Bochum, D-44801 Bochum, Germany}

\begin{abstract}
The discovery of superconductivity in infinite-layer nickelates has added a new family of materials to the fascinating  growing class of unconventional superconductors. By incorporating the strongly correlated multi-orbital nature of the low-energy electronic degrees of freedom, we compute the leading superconducting instability from magnetic fluctuations relevant for infinite-layer nickelates. Specifically, by properly including the doping dependence of the Ni $d_{x^2-y^2}$ and $d_{z^2}$ orbitals as well as the self-doping band, we uncover a transition from $d$-wave pairing symmetry to nodal $s_\pm$ superconductivity, driven by strong fluctuations in the $d_{z^2}$-dominated orbital states. We discuss the properties of the resulting superconducting condensates in light of recent tunneling and penetration depth experiments probing the detailed superconducting gap structure of these materials.  
\end{abstract}

\maketitle

\textit{Introduction.---}
Unconventional superconductivity is a hallmark signature of strongly
correlated materials. Electron pairing in heavy-fermion systems\;\cite{steg79} and high-$T_{\rm c}$ cuprates\;\cite{bed86} are only the most prominent examples thereof.
The finding of a superconducting (SC) state in rare-earth based
layered Ni$(3d^{9-\delta})$ oxides\;\cite{li19} adds another chapter to the
fascinating story. Key focus is on hole-doped infinite-layer
nickelates\;\cite{li19,zen20,osa20-1}, and a SC
multilayer system has also been discovered\;\cite{pan21}.
Right away it was thought that, because of their similarity to  Cu$(3d^9)$ oxides, the long sought-after 'cuprate-akin' pairing of $d_{x^2-y^2}$-wave symmetry had been finally found in nickelates. But follow-up studies cast some doubt on this view. Two distinct types of gaps were detected in tunneling measurements, one of which exhibits a full gap\;\cite{gu20}, and the isotropic character of the upper critical field $H_{\rm c2}$\;\cite{xia20,Wang2021} is also very different from cuprates. Furthermore, recently an analysis of the London penetration depth well below $T_{\rm c}$ seems incompatible with a pure $d_{x^2-y^2}$ paired SC state\;\cite{Chow_2022,Harvey_2022} and further $H_{\rm c2}$ measurements unveiled Pauli-limit violation\cite{Chow_2204_12606}.

At the normal-state level, several important differences between the correlated electronic structure of cuprates and infinite-layer nickelates have been identified. In the nickelates, there is no clear evidence for long-range antiferromagnetic (AFM) order\;\cite{hay99,wang20,Fowlie2022}, the doping-dependent Hall data point to a two-band scenario, and the SC dome is sandwiched between weakly-insulating doping regions\;\cite{li20,zen20}. Moreover, the nickelate charge-transfer energy is larger\;\cite{jia19,lec20-1,shen2021role}, pointing to a competing Mott-Hubbard vs. charge-transfer regime. In line with this, weaker Zhang-Rice\;\cite{zha88} physics and possible multi-orbital processes have been revealed from electron energy-loss spectroscopy\;\cite{goo21}.

Calculations based on density functional theory (DFT) uncovered, besides a dominant Ni-$d_{x^2-y^2}$ dispersion, the importance of self-doping (SD) bands stemming from hybridizing Ni$(3d)$ and rare-earth $(5d)$ orbitals\;\cite{wu19,nom19,bot20,lec20-1}. When including electronic correlations beyond DFT, the precise role of the Ni degrees
of freedom and their mutual interplay with the SD character remains controversial. The main debate is between moderately-to-strongly correlated Ni-$d_{x^2-y^2}$-driven physics, and multi-orbital Ni$(3d)$ mechanisms\;\cite{bot21,che22}. In the former case, SC
nickelates are described as cuprate-like with dominant $d_{x^2-y^2}$-wave
SC pairing\;\cite{wu19,saka20,kit20,Adhikary_2020,Congjun_Wu_2021,Xie_2021,Jiang_2022,karp2022}. If additional SD-driven Kondo physics is taken into account, other intriguing $d$- and $s$-wave pairing solutions may become stabilized\;\cite{wangzhang2020}. Furthermore, various multi-orbital SC scenarios have been suggested\;\cite{wer20,zhav20}, but in most cases those are based on rather simplified descriptions of the realistic correlated electronic structure. By contrast, approaching infinite-layer nickelates with a combination of DFT and dynamical-mean field theory (DMFT), including the effect of explicit
Coulomb interactions on oxygen ions via self-interaction correction (SIC), in fact encourages the Ni multiorbital viewpoint for the normal state. In such DFT+sicDMFT calculations\;\cite{lec20-1,lec20-2,lec21-1}, the SC doping region is ruled by a strong interplay of Ni-$e_g$ \{$d_{z^2}$,$d_{x^2-y^2}$\} degrees of freedom. Relevant Ni-$e_g$ physics is also suggested from Refs.\;\cite{Wan_2021,Kang_2021}.

Here, motivated by the recent experimental evidence against a simple SC $d_{x^2-y^2}$-wave scenario, we perform a detailed theoretical investigation of the leading pairing instabilities within a realistic multiorbital three-dimensional (3D) model for infinite-layer nickelates. The pairing kernel is generated by magnetic fluctuations, which are constrained by recent resonant inelastic x-ray scattering (RIXS) measurements mapping out dispersive magnetic modes in these materials\;\cite{Lu_2021,Lin_PRL_2021}. Intriguingly, the modifications of the electronic bands caused by doping naturally induces a transition from $d$-wave order to $s_\pm$ SC. We uncover the microscopic origin of this transition in the leading pairing symmetry, and discuss consequences for experiments probing the SC spectral gap.    

\begin{figure}[tb]
\includegraphics[width=\linewidth]{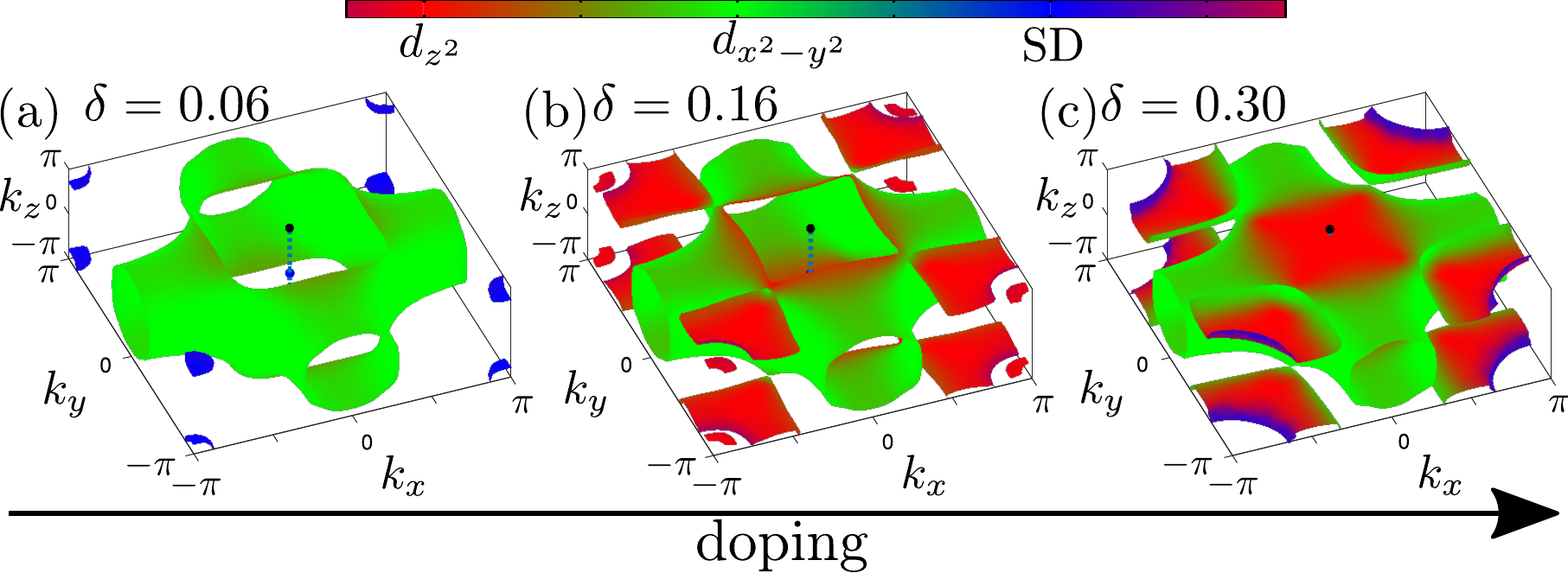}
\caption{Interacting Fermi surface (FS) with increasing hole doping $\delta$. (a) At $\delta=0.06$, the FS is dominated by the $d_{x^2-y^2}$ orbital (green). (b) Upon increasing $\delta$, the pockets at $(\pi,\pi,\pi)$ [all momenta in units of the respective inverse lattice constants] become sizeable and nest well with the larger FS sheet that is also of $d_{z^2}$ character (red) (c). In all cases, the SD orbital (blue) has small contributions on the Fermi level.}\label{fig_fs}
\end{figure}
\textit{Electronic model.---} Treating correlation effects emerging from the transition-metal (TM) {\sl and} the ligand oxygen sites on equal footing proves important for late TM oxides with significant competition between Mott-Hubbard and charge-transfer physics.
As a result, the DFT+sicDMFT approach\;\cite{lec19} to infinite-layer nickelates, with their anisotropic TM$(3d)$-O$(2p)$ bonding situation due to the lack of apical oxygen, gives rise to a dichotomic Mott-critical regime of (nearly-)insulating Ni-$d_{x^2-y^2}$ and itinerant Ni-$d_{z^2}$ orbital states. 

This picture remains valid in a rotational-invariant slave-boson\;\cite{li89,lec07} assessment of an effective three-band model, tailored to reproduce the key DFT+sicDMFT findings at stoichiometry\;\cite{lec20-2}. The model hopping integrals $t_{ij}^{\ell \ell'}$ are derived from a Wannier downfolding of the DFT band structure; including Ni-$d_{z^2}$, Ni-$d_{x^2-y^2}$ and the SD band. Supplemented with local Coulomb interactions, the effective Hamiltonian reads\;\cite{lec20-2} 
\begin{equation}                                    H=\sum_{i\ne j,\ell \ell',\sigma} t_{ij}^{\ell \ell'}
\,c^\dagger_{i\ell\sigma} c^{\hfill}_{j\ell'\sigma}+\sum_i\left( H^{(i)}_{\rm int} + H^{(i)}_{\rm orb}\right),\label{eq_ham}   \end{equation}
for $\ell,\ell'=d_{z^2},d_{x^2-y^2},{\rm SD}$ and lattice sites $i,j$. Note that while the nearest-neighbor (NN) hopping between
$d_{z^2}$ and SD is sizable with $\sim 90$\,meV, it is absent between $d_{x^2-y^2}$ and SD orbitals. Moreover, the in-plane NN hopping within $d_{x^2-y^2}$ is substantial with $t_x\sim{}\;390$\,meV, while within $d_{z^2}$ a comparable NN $t_z\sim 400$\,meV along the
$c$-axis is active. This highlights the competition of 2D vs. 3D characteristics. The onsite interaction part $H^{(i)}_{\rm int}$, notably only for the Ni-$e_g$ orbitals, has a two-orbital Slater-Kanamori form, i.e. includes density-density terms as well as pair-hopping and spin-flip terms, and is parameterized by Hubbard $U$ and Hund's coupling $J$. 
The remaining non-interacting onsite $H^{(i)}_{\rm orb}$ carries crystal-field terms via the onsite levels $\varepsilon_\ell$, a double-counting correction in the fully-localized-limit form\;\cite{ani93} as well as a potential-shift term for the SD orbital. The latter proves necessary to keep the SD band at the stoichiometric Fermi level, in line with the DFT+sicDMFT result\;\cite{lec20-2}. Model (\ref{eq_ham}) describes a (near) orbital-selective Mott transition for $d_{x^2-y^2}$ at $U_c=7$\,eV, $J=1$\,eV. Whilst the oxygen degrees of freedom are integrated out, the model thus restores the DFT+sicDMFT picture\cite{lec20-1}, since the realistic interplay of Ni- and O-based correlations leads to the identical highly correlated regime. The impact of the O-based correlations is hence taken into account properly in the model, and carried over to finite hole doping. The key correlation effects from the
slave-boson solution are then twofold: renormalization of the dispersion via the quasiparticle (QP) weights $Z_{\ell\ell'}^{\text{sb}}$ {\sl and} renormalization of the effective onsite levels via the shifts $\Delta_{\ell}=\tilde{\varepsilon}_\ell-\varepsilon_\ell$, with $\tilde{\varepsilon}_\ell$ as the interacting level energy. With doping, the non-zero quantities $\Delta_\ell$ shift the $d_{z^2}$-dominated flat band across the Fermi level, resulting in an additional $d_{z^2}$-dominated Fermi-surface sheet (see Fig.\;\ref{fig_fs}).
Both effects can be parametrized in a fully renormalized band structure with the Hamiltonian
\begin{equation}
    H=\sum_{i, j,\ell \ell',\sigma} \tilde t_{ij}^{\ell\ell'}\,c^\dagger_{i\ell\sigma} c^{\hfill}_{j\ell'\sigma},\label{eq_ham_ren}
\end{equation}
with renormalized hoppings $\tilde t_{ij}^{\ell\ell'}$ taken from Ref.\;\onlinecite{lec20-2}, which is the basis of all further analysis in this work.

\begin{figure}[b]
\includegraphics[width=\linewidth]{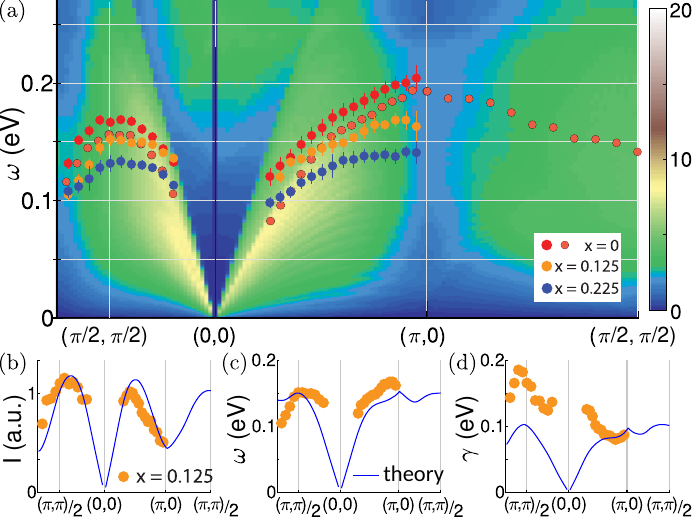}
\caption{Paramagnetic RIXS response. (a) Calculated RIXS intensity (map) in comparison with experimental data points as presented in Figs. 2A and 4E of Ref.\;\onlinecite{Lu_2021}. The calculation uses the correlated dispersion from Fig.\;\ref{fig_fs}(b), and setup parameters (Brillouin zone cut, scattering geometry) from experiment. Interactions on the Ni $d$ orbitals read $\tilde U=0.8 \tilde U_c$, $\tilde J=\tilde U/5$, for a fully coherent electronic structure. (b-d) Parameters for the weight $I$ (adjusted a.u.), position of the maximum $\omega$ and broadening $\gamma$ from a fit to a Lorenzian compared to experimental data from Fig. 4D,E,F of Ref.\;\onlinecite{Lu_2021}.}\label{fig_chi}
\end{figure}

\begin{figure*}[t]
\includegraphics[width=\linewidth]{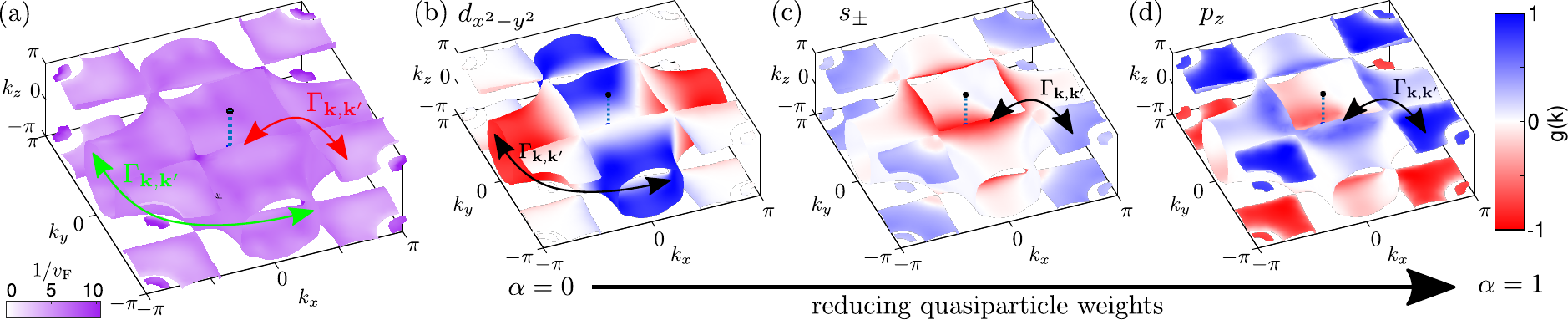}
\caption{Pairing structure with correlations. (a) The FS exhibits weak Fermi-velocity $v_{\mathrm F}$ anisotropy implying that the pairing interaction $\Gamma_{\bf k, \bf k'}$ in both the $d_{x^2-y^2}$ (green) and $d_{z^2}$ (red) orbitals is  active. (b-c) Effect of correlations on the gap symmetry function $g({\bf k})$ of the leading instability. Upon reducing the QP weights $Z_\ell$ by tuning $\alpha$, a $d_{x^2-y^2}$ pairing instability (b) transitions into a sign-changing $s_\pm$ state (c), and eventually into a spin-triplet $p_z$ state (d). For (c,d) the dominant pairing interactions are in the $d_{z^2}$ orbital states. Black arrows in (b-d): dominating pairing interactions (projected to band space).
Used parameters are $\tilde U=0.8U_c$ and $\tilde J/\tilde U=1/5$.} 
\label{fig_mechanism}
\end{figure*}

\textit{Magnetic fluctuations.---} Intrinsic magnetism and substantial magnetic fluctuations have been recently revealed by several experimental probes\;\cite{Lu_2021,Krieger_2021,Fowlie2022}, and have also been identified as important from first-principles calculations \cite{lec21-1,Leonov_2020,Leonov_2021}. This motivates further theoretical studies both of these magnetic fluctuations themselves, as well as their implication for pairing. Specifically, resonant inelastic x-ray scattering (RIXS) measurements on the Ni L$_3$-edge have uncovered a dispersive paramagnon-like mode emanating from the $\Gamma$-point with a bandwidth of approximately 200\;meV, see Fig.\;\ref{fig_chi}\;\cite{Lu_2021,Krieger_2021}.
Starting from the slave-boson solution to the effective Hamiltonian (Eq.\;(\ref{eq_ham_ren})), we follow the same procedure as in Refs. \onlinecite{Kaneshita_2011,Kreisel_RIXS} to compute the RIXS intensity. The RIXS process consists of a resonant excitation of a Ni $p$ state and a subsequent relaxation of a $d$ state together with the emission of a photon that is detected.
The calculation in second-order perturbation theory leads to a sum of elements of the susceptibility tensor weighted with the dipole matrix elements connecting Ni $d$ and $p$ states, taking into account the polarization vectors of incoming and outgoing x-rays\;\cite{Kreisel_RIXS}.
The matrix elements are obtained from atomic-like orbitals and we calculate the bare susceptibility tensor $\chi^0_{\ell_1 \ell_2 \ell_3 \ell_4}({\bf q},\omega)$ with $\ell_i$ denoting orbitals from the multiband model, Eq.\;(\ref{eq_ham_ren}). We employ the random phase approximation (RPA) with renormalized interaction parameters $\tilde U$, $\tilde J$ for the paramagnetic state\;\cite{Kreisel_RIXS}.
As demonstrated in Fig.\;\ref{fig_chi} (b-d), the experimentally detected paramagnon mode is almost quantitatively reproduced with exception of the tendency of smaller overall weight and less broadening towards $(\pi,\pi)$ where however the RIXS measurements are close to the limit of kinematic constraints. Since magnetic fluctuations are well-known to be capable of mediating Cooper pairing\;\cite{Scalapino_RMP,Romer2015}, a natural next question is what kind of SC condensate they support in infinite-layer nickelates?  

\textit{Superconductivity.---} Spin-fluctuation mediated pairing has become a standard framework for determining the order parameter symmetry and the associated momentum structure of the SC gap function in systems prone to magnetic ordering\;\cite{Scalapino_RMP,Kreisel_review}. The spin-fluctuation approach has been shown to be well-aligned with several brute-force numerical methods\;\cite{Romer_prr} and has e.g. successfully described $d_{x^2-y^2}$-wave pairing in cuprates and $s_\pm$ symmetry in SC iron pnictides. Remarkably, for multiorbital systems with strong orbital differentiation near the Fermi level, a simple orbital-selective generalization of this approach seems to apply\;\cite{Sprau2017,Kreisel2017,Romer_prl_2019,Kreisel_review,Bjornson_2020}. The method includes reduced pairing tendency from the most strongly-correlated states, as these exhibit reduced QP weights at low energy. We stress that the self-consistently renormalized pair vertex is not incorporated, and the scheme merely mimics the effects of reduced orbital-dependent QP weights on the pairing structure\;\cite{Bjornson_2020}. Approximating those weights by the diagonal part of the slave-boson QP-weight matrix, i.e. $Z_{\ell\ell'}=\delta_{\ell,\ell'}Z_\ell$, the bare susceptibility acquires prefactors and reads $\sqrt{Z_{\ell_1} Z_{\ell_2}Z_{\ell_3}Z_{\ell_4}}\chi^0_{\ell_1 \ell_2 \ell_3 \ell_4}({\bf q},\omega)$. Within RPA this leads to the renormalized pairing interaction $\Gamma_{\bf k, \bf k'}$ when projected to the eigenstates at $\bf k$, $\bf k'$ on the FS\;\cite{Sprau2017,Kreisel2017}. Solution of the associated linearized gap equation
 \begin{equation}\label{eqn:gapeqn}
-\frac{1}{V_G}
\sum_\mu\int_{\text{FS}_\mu}dS'\; \Gamma_{\bf k,\bf k'} \frac{g_i(\bf k')}{|v_{\text{F}\mu}(\bf k')|}=\lambda_i g_{i}(\bf k)\,,
 \end{equation}
yields the gap symmetry function $g_i({\mathbf k})$ of the instability $i$ with eigenvalue $\lambda_i$\;\cite{Romer_prl_2019,Kreisel2017}. The sum $\mu$ runs over the bands crossing the Fermi level, the integral is over the FS points $\bf k'$, $V_G$ is the Brillouin zone volume and $v_{\text{F}\mu}(\bf k')$ denotes the Fermi velocity at point $\bf k'$ as obtained from the correlated model\;(\ref{eq_ham_ren}).

The DFT+sicDMFT calculations\;\cite{lec20-1,lec20-2,lec21-1} deduced that the correlated $d_{x^2-y^2}$ and $d_{z^2}$ Ni orbitals have significant weight at low energies, and hence contribute to SC pairing in a spin-fluctuation mechanism. After the SD-band eliminating Lifshitz transition, which is also discussed in Refs.\;\cite{Wang_2020,Leonov_2020}, the low-energy spectral weight of the $d_{z^2}$ orbital increases as a function of hole doping, and yields a well-nested FS sheet (flat red area in Fig.\;\ref{fig_fs}(c)). In order to incorporate the effects of reduced coherence, we include the QP weights as obtained from the slave-boson approach into the computation of the SC instability. This changes the magnetic spectrum from being dominated by $(\pi,\pi,0)$ fluctuations at low doping $\delta=0.06$, to a case where significant contributions with momentum transfer at $(\pi,0,q_z)$ are present. To understand the effects of correlations on the SC instability within a simple picture, it is sufficient to consider the intra-orbital pair scattering between FS points $\bf k$, $\bf k'$ with same orbital component $\ell$ (same color in Fig.\;\ref{fig_fs}). In the modified spin-fluctuation approach, one finds a proportionality to the (square) of the QP weight, $\Gamma_{\bf k,\bf k'}\propto Z_\ell^2\chi^0_{\ell \ell \ell \ell}({\bf k}-{\bf k}',0)$. Therefore, reducing $Z_l$ gradually switches off the contribution of that orbital component in the gap equation, Eq.\;(\ref{eqn:gapeqn}), in favor of the orbitals which retain sizable $Z_\ell$. The relative contribution of the $d_{z^2}$ susceptibility then increases as also observed in a previous DMFT study \cite{Leonov_2020}. In this manner the reduced QP weights lead to a weakening of the $d_{x^2-y^2}$-wave pairing instability, as also discussed in view of increasing weight of the $d_{z^2}$ orbital in cuprates \cite{Sakakibara_2010}, eventually preferring other symmetry-allowed solutions with sign change along $k_z$ caused by the strongly-corrugated FS.

To illustrate this transition in the preferred pairing symmetry, we concentrate on the doping level $\delta=0.16$, and display in Fig.\;\ref{fig_mechanism}(a) the inverse Fermi velocity. It is essentially constant over the entire area, i.e. pair scattering processes involving all $k_F$ momenta are relevant. The dominant pair fluctuations in the $d_{z^2}$ orbital channel (red) and in the $d_{x^2-y^2}$ orbital channel (green) are highlighted. In the absence of orbital-selective QP-weight renormalization in the pairing kernel, i.e. starting from a model with $Z_\ell=1$ for all orbitals $\ell$, Cooper pairing from the $d_{x^2-y^2}$ orbital strongly dominates, leading to $d_{x^2-y^2}$-wave order (cf. Fig.\;\ref{fig_mechanism}(b)). This SC state is similar to the cuprate one, with small deviations due to the 3D nature of the FS, and agrees with previous theoretical studies of spin-fluctuation-like approaches to pairing in infinite-layer nickelates\;\cite{wu19,saka20,kit20,Adhikary_2020,Congjun_Wu_2021,karp2022}, for a comparison see the Supplemental Material\;\cite{Note1}.
But increasing effects of correlations in the pairing and thereby proportionally reducing the respective QP weight $\sqrt{Z_\ell}$ for orbital $\ell$ to $\alpha(\sqrt{Z_\ell^\text{sb}}-1)+1$ where $\alpha \in [0...1]$, the $d_{z^2}$ orbital maintains significant coherence at low energies because it is not close to a Mott-critical regime. This results in a leading $s_\pm$-wave pairing instability displayed in Fig.\;\ref{fig_mechanism}(c) with sign changes between the red parts of the inner FS sheet and the pockets close to $(\pi,\pi,\pi)$, driven by the $(\pi,0,q_z)$ magnetic fluctuations. Importantly, this state remains nodal with vanishing gap on the large FS sheet, see Fig.\;\ref{fig_mechanism}(c)
When reducing QP weights further, $\alpha\rightarrow 1$, i.e. $Z_\ell=Z_{\ell}^{\text{sb}}$, surprisingly, the dominant instability is an odd-parity spin-triplet nodal SC state with $p_z$ symmetry (see Fig.\;\ref{fig_mechanism}(d)), exhibiting again a full gap on the $d_{z^2}$ dominated FS. At present, there are controversial reports on Pauli-limited superconductivity\;\cite{Wang2021,Chow_2204_12606} such that a triplet state may be realized in infinite-layer nickelates. Finally, we note an interesting subleading even-parity two-dimensional $E_g$: $d_{xz}$/$d_{yz}$ state is close by in energy, but does not become leading for the parameter regimes we have explored in this work. Additional discussion on this state, the parameter dependence of our results and stability of the triplet state are presented in the Supplemental Material\;\cite{Note1}.

 \begin{figure}[t]
\includegraphics[width=\linewidth]{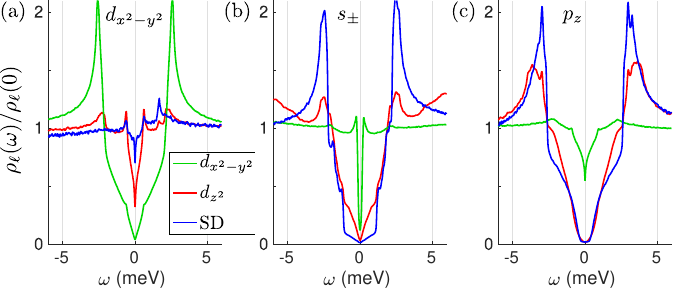}
\caption{Orbitally-resolved density of states $\rho_\ell(\omega)$ for the three SC phases presented in Fig.\;\ref{fig_mechanism}. (a) $d$-wave state from a fully coherent electronic structure featuring a large V-shaped gap on the $d_{x^2-y^2}$ orbital, (b) sign changing $s_\pm$ state with small gap in the largely incoherent $d_{x^2-y^2}$ orbital channel, and (c) $p_z$-wave spin-triplet state. The three curves correspond to the different orbital DOS components, normalized by their normal-state value at the Fermi level $\rho_\ell(0)$.}
\label{fig_dos}
\end{figure}
 
\textit{Superconducting spectral properties.---} Our important message is that incorporation of the 3D FS {\it and} the significant doping difference of the $d_{z^2}$ vs. $d_{x^2-y^2}$ orbital channels, leads to a transition in the fundamental pairing symmetry. Infinite-layer nickelates are thus distinct from cuprates, where $d_{x^2-y^2}$-superconductivity is the sovereign ruler. They are instead more comparable to e.g. iron pnictides\;\cite{Scalapino_RMP,Kreisel_review} or Sr$_2$RuO$_4$\;\cite{Romer_prl_2019,Firmo_2013}. This highlights the possibility of accidentally degenerate SC symmetries and the need for careful comparison to experiments in order to pinpoint the SC order parameter symmetry. 

Focusing on the low-energy spectral properties, Fig.\;\ref{fig_dos} shows the orbital-resolved density of states (DOS) $\rho_\ell(\omega)$ for the three cases displayed in Fig.\;\ref{fig_mechanism}(b-d). Analyzing $\rho_\ell(\omega)$ is crucial because the tunneling conductance in scanning-tunneling microscope (STM) experiments is strongly influenced by the nature of the surface atoms\;\cite{Choubey2017}. Nickelate STM data have revealed a fully-gapped ``U-shaped'' spectra on some surfaces, while ``V-shaped'' spectra, reminiscent of a nodal superconductor, are found on others\;\cite{gu20}. It has been shown theoretically that spectra can interpolate between the partial DOS of one or the other internal degree of freedom by variation of the tip position within the unit cell\;\cite{Choubey2021}. Similar effects are expected from a tip-sample distance variation or if other surface atoms are present. At the moment, an {\it ab initio} calculation using surface Wannier functions is out of reach since the details of the surface are unknown. However, examining $\rho_\ell(\omega)/\rho_\ell(0)$ for the $s_\pm$ state, Fig.\;\ref{fig_dos}(b), one observes a V-shaped nodal behavior for the $d_{z^2}$ orbital and U-shaped DOS in the SD orbital. The $d_{x^2-y^2}$ orbital has a very small gap and may not be tunneled into because of its in-plane structure. This should lead to a suppressed value of the wave functions toward the vacuum\;\cite{Firmo_2013,Kreisel2021}. By contrast, the SD orbital states are hybridized Wannier functions with significant weight distant to the NiO$_2$ plane\;\cite{{lec20-2}} and may be responsible for the STM-observed full-gap conductance (cf. Fig.\;\ref{fig_dos}(b))\;\cite{gu20}. The same argumentation for the d-wave SC order parameter would lead to the hardly gapped spectrum in Fig.\;\ref{fig_dos}(a) if the in-plane orbital component cannot be picked up by the STM tip. Finally, the low temperature $T$ behavior of the penetration depth $\lambda(T)$ can be used to quantify the QP excitation spectrum at low energies\;\cite{Chow_2022,Harvey_2022}. In the present case, the calculated nodal SC order parameters in a clean material lead to a linear dependence of $\lambda(T)$ at low $T$, as we have verified numerically. This seems consistent with recent experiments on optimally doped La- and Pr-nickelates\;\cite{Harvey_2022} when taking into account disorder effects\;\cite{Hirschfeld93}. For Nd-based nickelates, however, the situation appears more complex\;\cite{Chow_2022}.

\textit{Conclusions.---} Guided by recent experimental findings reporting significance of magnetism in infinite-layer nickelates, we have applied a microscopic model to obtain the pairing kernel from magnetic fluctuations relevant to these materials. In line with earlier studies, $d_{x^2-y^2}$ Cooper pairing is found as a prominent candidate SC state. However, as a function of enhanced electronic correlations, we have uncovered a tendency for these systems to transition from $d_{x^2-y^2}$ order to nodal $s_\pm$ superconductivity. This appears consistent with recent experiments finding evidence of partial full-gap spectral properties of the SC state of infinite-layer nickelates.  

\begin{acknowledgments}
\textit{Acknowledgements.---} We acknowledge useful discussions with F. Jakubczyk. We thank P. Buzduga for her contribution in the data analysis of the theoretical RIXS spectra. A. T. R.  and B. M. A. acknowledge support from the Independent Research Fund Denmark, grant number 8021-00047B. I.M.E. and F.L are supported by the German Research Foundation within the bilateral NSFC-DFG Project ER 463/14-1. 
\end{acknowledgments}

%
\ifarXiv
    

\section*{Supplementary material}

# Supplementary Material: Superconducting Instabilities in Strongly-Correlated Infinite-Layer Nickelates

Andreas Kreisel,[1] Brian M. Andersen,[2] Astrid T. Rømer,[2] Ilya M. Eremin,[3] and Frank Lechermann[3]

[1]*Institut für Theoretische Physik, Universität Leipzig, D-04103 Leipzig, Germany*
[2]*Niels Bohr Institute, University of Copenhagen, 2100 Copenhagen, Denmark*
[3]*Institut für Theoretische Physik III, Ruhr-Universität Bochum, D-44801 Bochum, Germany*

This supplementary material contains additional results of the pairing calculations presented in the main text. The multi-orbital nature in conjunction with correlations induce effects of orbital-differentiation leading to several competing instabilities, one of them a sign-changing $s_\pm$ order parameter.

## Pairing calculation from a bare electronic structure without correlations

In the main text, we present our approach to include electronic correlations in (a) the band structure itself which contains orbitally-dependent band renormalizations and onsite-shifts from the slave-boson approach mimicking the results of DFT+sicDMFT calculations [2–4] and (b) correlations in the pairing calculation itself from the reduced coherence of the low-energy quasiparticles as employed via the quasiparticle weights $Z_l$[5]. We show in Fig. 3 (b) of the main text the dominating $d$-wave instability when only including (a). To directly connect to earlier works on that subject, we show in Fig. S1 a corresponding result where no correlations are taken into account, i.e. where a three-band model as derived from the DFT approach only is used to calculate the superconducting instabilities. It turns out that for the doping level considered, $\delta = 0.16$, there is a sizeable $\Gamma$ centered electron-like pocket which barely participates in pairing. Instead, we find a dominant $d_{x^2-y^2}$ instability (panel b) driven by the $d_{x^2-y^2}$ orbital component, green in panel (a), followed by a $g$-wave and another $d_{x^2-y^2}$ pairing instability.

## Parameter dependence and details of pairing states in the correlated regime

To tune the correlations in the pairing calculation, we scale the (root) of the quasiparticle weight according to $\sqrt{Z_\ell} = \alpha(\sqrt{Z_\ell^{\text{sb}}} - 1) + 1$ where $Z_\ell^{\text{sb}}$ is the quasiparticle weight as obtained from the slave-boson approach. This yields the evolution of the quasiparticle weights presented in Fig. S2. This produces a transition from the dominating $d_{x^2-y^2}$ pairing state at $\alpha = 0$ to an $s_\pm$ pairing state with several near degeneracies, and finally into a triplet pairing state of $p_z$ symmetry at $\alpha = 1$. Details on the dependence of the pairing instability on the choice of the overall Hubbard interaction $U/U_c$ in units of the critical $U_c$ for reaching the magnetic instability in RPA and the choice of $J/U$ are shown in Fig. S3: in panel (a), the clear dominance of the $d_{x^2-y^2}$ pairing state for all choices of $U$ is visible at $\alpha = 0$, while in panel (b) at $\alpha = 0.5$, several pairing states such as $s_\pm$, $d_{xz,yz}$ or the $p_z$ triplet solution are competing with moderate dependence on the overall scale $U/U_c$ (different lines cross) and weak dependence on the choice of $J/U$ (lines of same color, but different linestyle, see legend). For $\alpha = 1$, there is again a clear separation of the instabilities, panel (c).

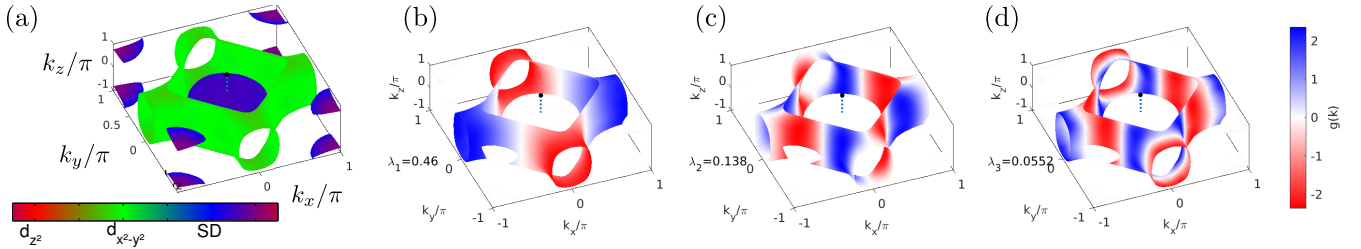

FIG. S1. Calculation at $\delta = 0.16$, but using the bare electronic structure, i.e. without effects on correlations in the bands and without correlations in the pairing. (a) Fermi surface (color reflects the orbital content). Three sheets are present: A $\Gamma$ centered (almost spherical pocket) from the self-doping orbital, a pocket centered at $(\pi, \pi, \pi)$ and the cuprate-like pocket dominated by the $d_{x^2-y^2}$ orbital (green). (b-d) Leading pairing instabilities from solutions of the linearized gap equation, similar as worked out in Ref. [1] where a $d_{x^2-y^2}$ solution is by far dominating (largest eigenvalue $\lambda$) followed by a $g$-wave state and a higher order $d_{x^2-y^2}$.

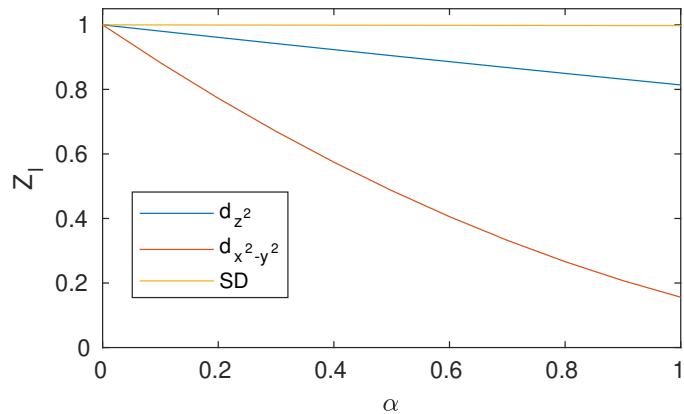

FIG. S2. Evolution of the quasiparticle weights $Z_l$ as a function of the parameter $\alpha$ describing the effects of correlations taken into account when calculating the pairing instability from a modified spin-fluctuation approach[5] including quasiparticle weights.

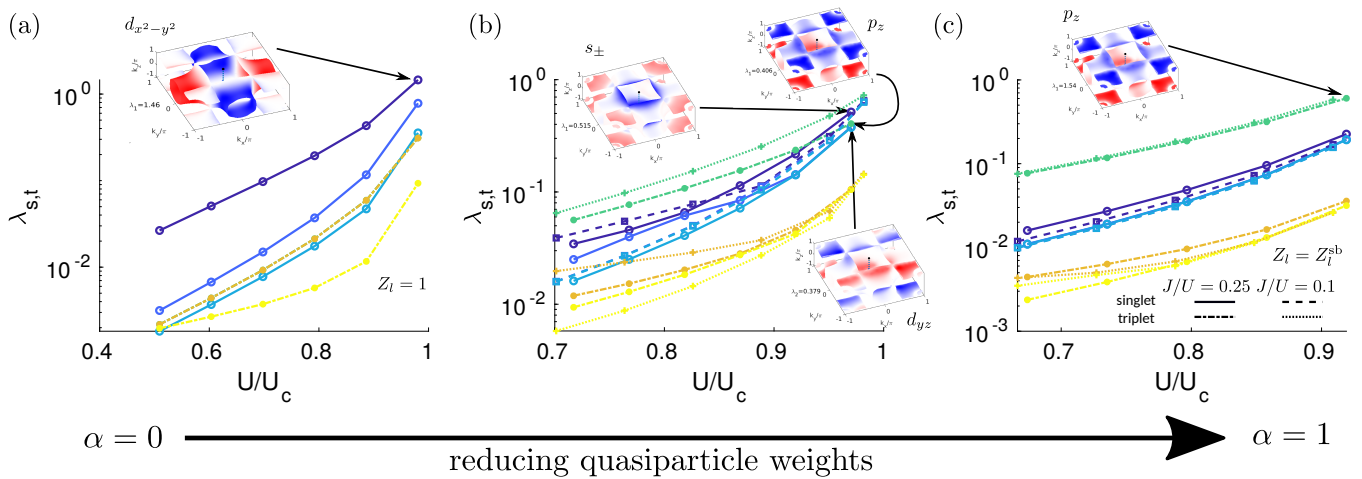

FIG. S3. Parameter dependence of the leading pairing states. Eigenvalues (different colors refer to first, second and third eigenvalue in singlet/triplet channel) of the linearized gap equation as a function of the interaction parameters $U$ (and $J$) when investigating the effects of correlations. (a) Uncorrelated electronic structure with the robust $d$-wave instability (inset) and almost no dependence on the ratio $J/U$ (full (dash-dotted) lines $J/U = 0.25$, dashed (dottet) lines $J/U = 0.1$ for singlet (triplet)). (b) Moderately correlated case where multiple pairing instabilities become close by and cross ($d$-wave, $s$-wave and triplet state in the inset with very similar eigenvalues. (c) Using the quasiparticle weights from the slave-boson calculation, the pairing solutions are well separated, and no crossings occur as a function of $U$.

Finally, in Fig. S4 the evolution of the leading eigenvalues as function of $\alpha$ is presented. We fix the bare interactions at $U/U_c = 0.96$ to reach sizeable eigenvalues in all regimes. Again, one observes the strong $d_{x^2-y^2}$ instability for the uncorrelated case $\alpha = 0$, as discussed in Fig. S1 and Fig. S3 (a). As correlations in the pairing calculation are gradually included, this instability gets suppressed due to the small $Z_{x^2-y^2}$ and eventually the sign-changing $s$-wave solution becomes the dominant singlet instability (followed by $d_{xz,yz}$ solutions, see Fig. S3 (b). Towards $\alpha = 1$, the quasiparticle weights for the $d_{x^2-y^2}$ orbital and for the $d_{z^2}$ orbital become small (see Fig. S2), and a triplet instability of symmetry $p_z$ acquires sizeable eigenvalues. Noting that its eigenvalues quickly surpass the singlet instabilities, we point out that in general, a triplet pairing acquires stronger frequency-dependent corrections when treated in an Eliashberg formalism such that a relative suppression of the triplet instability compared to the singlet ones is expected[6] and the overall range in $\alpha$ where a singlet instability is leading should be extended.

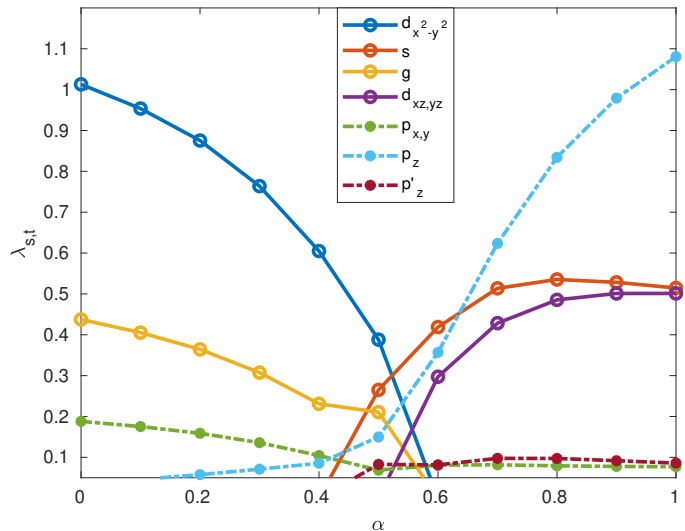

FIG. S4. Eigenvalues as a function of the parameter $\alpha$ as defined in the manuscript for a given ratio $U/U_c = 0.96$ (to achieve sizable pairing eigenvalues $\lambda \approx 1$). One can see a clear trend that the eigenvalue of the $d_{x^2-y^2}$ solution (dark blue) rapidly decreases as correlations are included in the pairing calculation, becoming practicably irrelevant at $\alpha > 0.5$. At the same time, the eigenvalue of the $s$-wave solution (red) increases and becomes the dominating singlet solution (before being overtaken by the $p_z$-type triplet solution).


\fi
\end{document}